# The Origin and Distance of the High-Velocity Cloud MI

J.T. Schmelz
0000-0001-7908-6940
USRA, 7178 Columbia Gateway Drive, Columbia, MD 21046
jschmelz@usra.edu

G.L. Verschuur
0000-0002-6160-1040
verschuur@aol.com

**Abstract**

The high-velocity, neutral hydrogen feature known as MI may be the result of a supernova that took place about 100,000 years ago at a distance of 163 pc. Low-velocity HI data show a clear cavity, a structure indicative of regions evacuated by old exploding stars, centered on the spatial coordinates of MI, (l,b) = (165°, 65.°5). The cavity is also visible in 100 micron dust data from IRAS. The invisible companion of the yellow giant star, 56 Ursae Majoris, may be the remains of the supernova that evacuated the cavity and blasted MI itself outward at 120 km/s. The mass and energy of MI are easily in line with what is expected from a supernova. The diffuse X-rays seen by ROSAT provide evidence of a hot cavity, and the enhanced X-rays may result from the subsequent bow shock.

**Introduction**

Of all the possible scenarios for the origin and energy of HI high-velocity (HV) clouds proposed by Oort (1966), perhaps the most obvious candidate - an old, local supernova - was considered and almost immediately rejected. Interestingly, he dismissed this hypothesis on what he considered as *rather convincing grounds*. The three main reasons for rejection were: (1) the artificial arrangement of the required nearby supernovae to explain all anomalous-velocity[1] gas seemed unlikely; (2) optical emission nebulosities, similar to those that define the Cygnus loop, were absent; and (3) the evidence from optical absorption studies that placed the HV features at kiloparsec distances was growing, making the energy considerations unlikely.

Traditional studies have indeed placed HV gas complexes at great distances, in the halo and beyond (see, e.g., van Woerden & Wakker 2004). This is certainly true for the Magellanic Stream, but distance estimates for other complexes are not as clear cut.[2] In principle, a star behind an HI cloud might show HV absorption features, but a star in front would not. If the distances to the stars are known, the cloud should lie between them (see, e.g., Kuntz & Danly 1996; Wakker 2001).

This seemingly straightforward method is complicated by the structure of the HI gas itself. Early observations found that diffuse interstellar HI gas can be filamentary (e.g., Verschuur 1973;

---

[1] Greater than about ±30 km/s at high Galactic latitudes
[2] Although differential rotation can be used to estimate distances to HI gas in the Galactic disk, these effects are negligible at high latitudes

Colomb et al. 1980). Verschuur (1991) showed that filamentary features twisting into and out of the line-of-sight could produce patches of enhanced emission that create the illusion of clouds. Recent surveys have revealed that the filamentary nature of HI is pervasive (McClure-Griffiths et al. 2009; Peek et al. 2011; Winkel et al. 2016; Martin et al. 2015). Thus, the assumption of homogeneity over the entire radio beam for a feature with a cloud-like appearance do not seem to hold up. In fact, Mebold et al. (1990) and Ryans et al. (1997) have warned against using a non-detection to constrain the distance, so the traditional absorption-line studies can provide a well-defined upper limit but not a lower limit to the distance.

In this paper, we revisit the supernova model rejected by Oort (1966) for the bright HV feature discovered by Mathewson (1967) and later designated as MI. We address the three main criticisms as follows: (1) we limit our investigation to explain the origin and energy of MI alone and resist the temptation to apply this scenario to other anomalous velocity features where the evidence may be less convincing; (2) the supernova in question may have exploded of order 100,000 years ago, old enough that any associated optical nebulosity has faded; (3) although absorption-line studies can only constrain MI to a distance less than several kiloparsecs, our results indicate that it may be much closer with realistic energy considerations.

**Analysis**

The $\lambda$-21-cm Galactic neutral atomic hydrogen data from the Effelsberg-Bonn HI Survey (EBHIS) covers the northern sky with a declination greater than -5º and an angular resolution of 10.′8 (Winkel et al. 2016). Fig. 1 shows an area map of the HI brightness temperature as a function of galactic longitude (l) and latitude (b) at -14 km/s. As expected, there is a lot of complex low-velocity structure, but what was not expected is the absence of emission – a cavity – centered at about (l,b) = (165º, 65.º5). These are also the spatial coordinates of MI, the HV feature that the refereed literature usually places at the Galactic halo/disk boundary. Although the cavity itself is neither unique nor rare, its alignment with MI was unexpected because of the traditional view that low- and high-velocity gas are at vastly different distances.

Fig. 2 shows HI maps at different velocities centered on MI and the unexpected cavity. Panels (a) and (b) depict the changing structure of MI at -120 km/s and -110 km/s. Panels (c) and (d) show the rim of the MI cavity, higher-density emission that appears to be piled up on the lower right quadrant highlighted by the dashed circle. The lower panels, (e) and (f), feature the MI cavity itself, which is indicative of regions evacuated by old exploding stars.

If MI represents the gas moving toward us as a result of this explosion, then there might be gas moving away from us on the far side. Panels (a) and (b) of Fig. 3 show only the faintest indication of such gas, and a search at greater positive velocities is limited by low-level systematic noise in the EBHIS data. Panels (c) and (d) show the MI cavity emission from the Wisconsin H$\alpha$ Mapper (WHAM) survey data (Haffner et al. 2003) at velocities +10 and +20 km/s. These maps show small-scale structures located around the rim and inside of the MI cavity, but the velocity limit of the WHAM data prevents us from extending our search to greater positive velocities. The supernova explosion is pushing the complex, clumpy medium seen in Fig. 3. This is why the rim seen in the lower right quadrant of panels (c) and (d) of Fig. 2 is notable. Here, we get a close approximation to a more ideal case, where the combination of the snowplow effect (Spitzer 1978) and limb brightening combine to make this segment stand out. At other positions, however, the hydrogen shows the clumpy, non-uniform structure that seems to prevail in this region.

The rarity of anomalous positive HV gas could have been used by Oort (1966) as a fourth reason to reject the supernova hypothesis, but the strength of this rejection may be mitigated by the location of the supernova, which we discuss next.

The giant star known as 56 Ursae Majoris (56 UMa; HD 98839) is visible to the naked eye with an apparent magnitude of 5.03 and a spectral type of G7III. It has a radial velocity of 1 km/s and a distance of 163 pc (Ducati 2002; Pourbaix et al. 2004; van Leeuwen 2007). Its galactic coordinates are (l, b) = (164.°73, 65.°76), placing it near the center of MI. Interestingly, it is the visible component of a single-lined spectroscopic binary with an orbital period of about 45 years. The unseen companion had been identified traditionally as a white dwarf, but if it were a neutron star, it could be remains of the supernova explosion that evacuated the MI cavity and blasted MI itself outward at 120 km/s.

56 UMa itself has been tagged as a Barium (Ba) star in the literature (Pilachowski 1977), with an overabundance of Ba and other s-process elements. Since s-process nucleosynthesis is unlikely to occur in stars prior to the asymptotic-giant-branch phase, the best explanation for the enhanced Ba is contamination by mass transfer from a giant companion that has now evolved to become a white dwarf. Observations indicate that all Ba stars are in binary systems (McClure et al. 1980) and that some of these show the ultraviolet excess expected from a white dwarf (Dominy & Lambert 1983; Gray et al. 2011). Both of these discoveries add significant evidence to the prevailing hypothesis described above.

So why then do we think that the invisible companion of 56 UMa might be a neutron star rather than a white dwarf? First, the Ba-star classification for 56 UMa has always been *marginal* (see, e.g., Lu et al. 1991) or even *false* (Liang et al. 2003). Second, radial velocity observations spanning many years only recently covered the full orbital period of 45.1±0.3 years (Griffin 2008; Jorissen et al. 2019); the historical uncertainty in the basic orbital properties allowed for the possibility that the mass of the companion was consistent with that of a white dwarf. Mass estimates for the primary, however, are consistently greater than 3.5 or even 4 solar masses, challenging Ba-star formation theories (Gondoin 1999; Liang et al. 2003).

In addition, assumptions imposed on the orbital inclination to compute the mass of the companion were based on observations of other Ba-star systems, so if 56 UMa is not a Ba star, that mass determination could be unreliable, and the more accurate orbital determinations made possible with Gaia DR3 may lead to a different result (A. Jorissen, private communication).

Escorza et al. (2022) combine radial-velocity data with proper motions from Hipparcos and Gaia to derive a mass of 1.31±0.12 solar masses for the invisible companion of 56 UMa. This is high enough that the asymptotic giant branch progenitor of the presumed white dwarf would have generated no s-process elements (Karinkuzhi et al. 2018), essential for the classification of 56 UMa as a Ba star. Fortunately, there is a small mass interval where low-mass neutron stars and high-mass white dwarfs can overlap, and the invisible companion falls within it - above the lowest-known neutron-star mass (1.174 solar masses; Martinez et al. 2015), and below the super-Chandrasekhar limit white dwarfs.

Escorza et al. (2022) also determine abundances for 14 heavy elements, some produced predominantly by the s-process and others by the r-process. 56 UMa shows moderate enhancement in s-process elements, which is consistent with its traditional Ba-star classification, but also r-process enhancement making it the first hybrid (s+r) abundance star found with solar metallicity. They conclude that the invisible companion is a neutron star and determine that the repercussions of the supernova might be responsible for the unusual abundance profile. The blast from the explosion not only contaminated 56 UMa, but may have exposed the inner layers as the envelope of the star was partially stripped away (Hirai et al 2014).

The expelled gas from the companion that predates the supernova was almost certainly clumpy (Hamann et al. 2008; Smith 2014). MI itself may very well be a collection of these surviving clumps. The supernova scenario then gives us a definitive distance to MI. The implications of this are discussed in the next section, but one thing to note is that MI lies in the direction of the Local Chimney (Welsh et al. 1999), a low-density, high-latitude extension of the Local Bubble (Frisch 1981). The bubble and chimney were mapped in 3D in a series of papers (see Lallement et al. 2003 and references therein) using absorption characteristics of the interstellar NaI D-line doublet at 5890 Å. They targeted over 1000 stars lying within approximately 350 pc of the Sun as determined by parallax measurements made with the Hipparcos satellite. This is important because the far side of the supernova is blowing into this low-density environment, and the positive-velocity features may have completely dissipated.

One of the more challenging aspects of the observations is to explain the low-velocity HI gas seen toward the center of the MI cavity in panels (d) and (e) of Fig. 2. This could be just a chance overlap, but a far more interesting and satisfying scenario appears to come from 56 UMa itself. If this yellow giant star has a mass of 3.5-4 solar masses, then it must have passed through the instability strip[3] of the Hertzsprung-Russell diagram at least once during its evolution off the main sequence. Pulsations as well as the gravitational pull of its companion may have enabled 56 UMa to shed its outer layers into the MI cavity.

Fig. 4 shows close ups of (a) this low-velocity HI emission and (b) MI. 56 UMa has a proper motion [RA, dec] = [-37.42, -14.26] mas/yr, which equates to a transverse velocity of 22.8 km/s at a distance of 163 pc. This vector is shown in red, and appears to be along an extended tail of the HI emission. In addition, the enhanced X-ray emission from the ROSAT observations (Herbstmeier et al. 1995) are plotted in greyscale. This high-energy emission is sometimes attributed to non-thermal electron bremsstrahlung arising as the HV cloud slams into the galactic disk. However, since the disk-halo border is absent in the Local Chimney, a different explanation is required to explain the associated high-energy emission. Figure 4c shows the MI cavity in the 100 micron dust map from IRAS; like the gas, much of the dust has been pushed to the rim by the blast. If this structure had been created by a supernova, we might expect to detect evidence for a hot cavity. Figure 4d shows the diffuse ¼ keV X-rays observed with ROSAT (Herbstmeier et al.

---

[3] A narrow region in the Hertzsprung-Russell diagram that contains several different types of variables including Cepheid, RR Lyrae, W Virginis, and ZZ Ceti stars

1995) in greyscale superposed on the integrated hydrogen map. These diffuse X-rays are from the lower-left half of the cavity where the integrated HI emission is weakest.

**Discussion**

If the supernova of the invisible companion of 56 UMa led to the high velocity of MI and the other observational properties described in the last section, then the distance to the star of 163±7 pc gives us a means to calculate some of the relevant physical parameters. The radius of MI cavity in Fig. 1 is about 4°.5 or 13 pc. If we make the simplifying assumption that the expansion of the cavity is spherically symmetric and take the observed velocity of MI as the expansion velocity, then the supernova took place about 100,000 years ago, and MI itself is about 150 pc away.

The EBHIS data allow us to estimate the average HI column density for the various peaks of MI, which is found to be about $110 \pm 26 \times 10^{18}$ cm$^{-2}$. Making the standard assumption that the depth is about equal to the width of the sub-structures seen in MI (0.°5), the volume density is $31 \pm 16$ cm$^{-3}$, the mass is about three solar masses, and the energy is $3.1 \times 10^{47}$ ergs. Integrating over the full area of the cavity the total energy is estimated to be $1.5 \times 10^{49}$ ergs, well within the energy budget of a typical supernova ($10^{51}$ ergs). Similarly, for the low-velocity HI gas inside the MI cavity, the total column density is $45 \times 10^{18}$ cm$^{-2}$, the volume density is 11 cm$^{-3}$, and the mass is 1.2 solar masses. Shedding a solar mass seems to be a reasonable result of the passage of the yellow giant through the instability strip.

As the binary star system and its associated MI cavity are moving through space along the vector shown in Fig. 4, the blast wave from the supernova would be interacting with the low-density interstellar medium of the Local Chimney. The enhanced X-rays seen by ROSAT could originate from the resulting bow shock.

We could only explore this scenario for MI because 56 UMa survived the supernova explosion of its binary companion. The Local Chimney could be littered with the corpses of dead stars that produced other anomalous-velocity gas features, but these would be much harder to detect. We, quite deliberately, avoided applying this hypothesis more generally, so we should note that other explanations like the galactic-fountain (Shapiro & Field 1976; Bregman 1980) and supershell (Verschuur 1993) models are still very much in play for the larger-scale distribution of anomalous-velocity HI.

**Conclusions**

The HV neutral hydrogen feature know as MI may be the result of an old supernova at a distance of 163 pc. Here we summarize our proposed scenario.

The low-velocity EBHIS data show a clear cavity centered at (l,b) = (165°, 65.°5), the spatial coordinates of MI.

Like the gas, the 100 micron dust has been pushed to the rim of the MI cavity by the supernova blast.

Both EBHIS and WHAM data show only limited evidence for small-scale structures at positive velocities because the far side of the supernova is blowing into the low-density environment of the Local Chimney, and these features have completely dissipated.

The invisible companion of the yellow giant star known as 56 UMa is the remains of the supernova that evacuated the MI cavity and shot MI itself outward at 120 km/s.

The low-velocity HI gas seen toward the center of the MI cavity originated from 56 UMa itself as the yellow giant evolved off the main sequence and passed through the instability strip of the Hertzsprung-Russell diagram.

If we make the simplifying assumption that the expansion of the MI cavity is spherically symmetric and take the observed velocity of MI as the expansion velocity, then the supernova took place about 100,000 years ago, and MI itself is about 150 pc away.

MI has a mass of three solar masses and an energy of $3.1 \times 10^{47}$ ergs; this energy is easily in line with what is expected from a supernova.

The mass of the prominent low-velocity HI gas inside the MI cavity is about 1.2 solar masses.

The enhanced X-rays seen by ROSAT originate from the resulting bow shock.

The diffuse X-ray emission detected by ROSAT provides tangible evidence for the hot cavity.

If our prediction - that the companion of 56 UMa is a neutron star - proves true, then MI may be the first HV feature with both a well-determined distance and a clear-cut origin story.

## Acknowledgments

We are grateful to T. Dame for providing us with his MacFits software that seamlessly allows us to unravel data cubes, to B. Winkel for sending us the EBHIS data, and to N. Brickhouse, W.B. Burton, W. Reach, and K. Sembach for useful advice. We are also especially grateful to A. Jorissen and A. Escorza for sharing their expertise on 56 UMa.

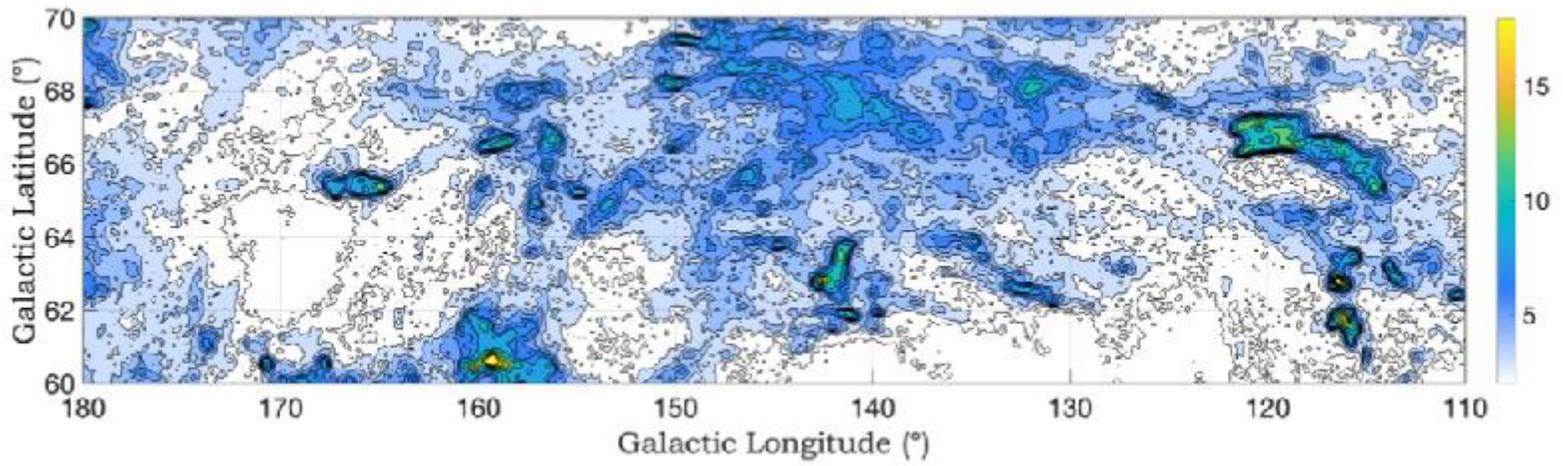

**Fig 1.** Area map at -14 km/s showing the complex low-velocity gas structure and the unexpected cavity between l = 158° and 175° using EBHIS data with a 2 km/s band width. Legend is in degrees K.

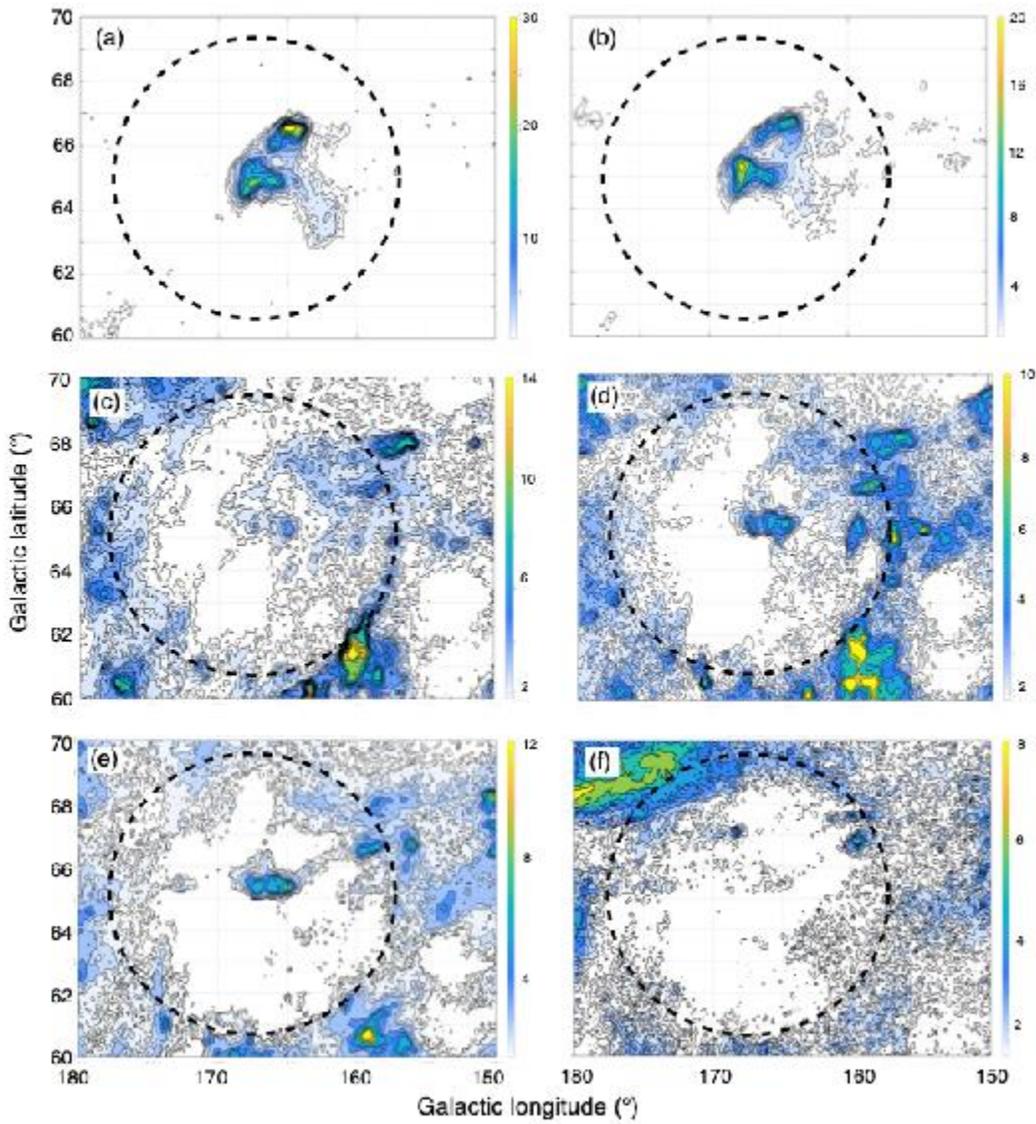

**Fig 2.** Area maps centered on MI showing the details of the environment at different velocities using EBHIS data with a 2 km/s band width. Legends are in degrees K. In each frame the dashed circle highlights the approximate boundary of the low-velocity cavity seen in Fig. 1. (a) the well-known two-peaked structure of MI at 120 km/s; (b) the structure of MI at -110 km/s; (c) the rim of a low-velocity cavity in the HI emission at -20 km/s, which is particularly prominent in the lower right quadrant; (d) the more pronounced rim at -16 km/s; (e) a low-velocity peak at the center of the cavity at -14 km/s; (f) the low-velocity cavity at 0 km/s.

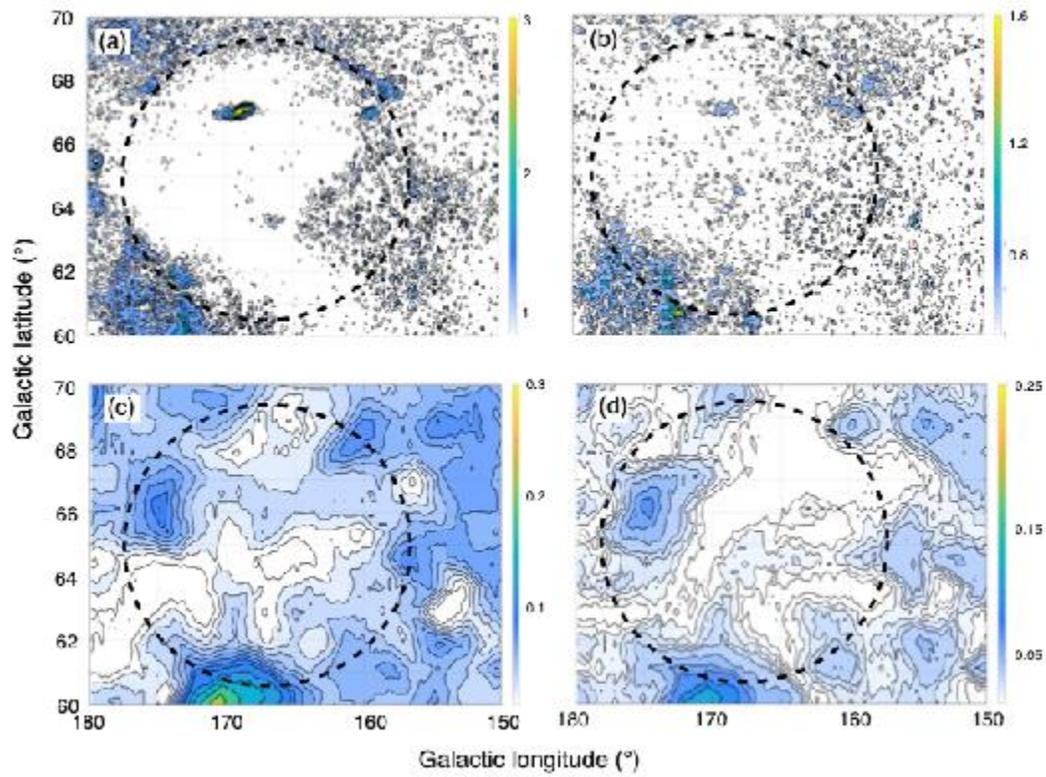

**Fig 3.** Positive velocity EBHIS HI maps at (a) +10 km/s and (b) +20 km/s with the legends in degrees K; and WHAM Hα maps integrated over 10 km/s intervals at (c) +10 km/s and (d) +20 km/s with the legends in Rayleighs. The dashed circle highlights the approximate boundary of the low-velocity cavity seen in Fig. 1.

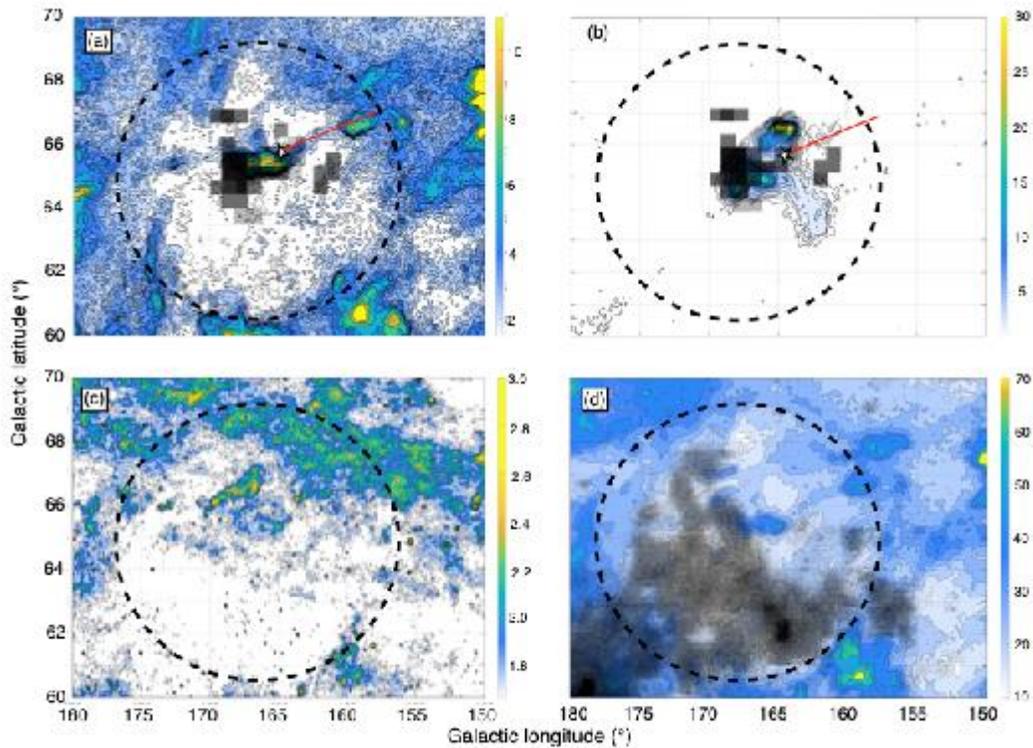

**Fig 4.** (a) Pixels showing the enhanced soft X-rays seen by ROSAT (greyscale; Herbstmeier et al. 1995), the current position of 56 UMa (star), and the proper motion vector (red arrow) overlaid on the low-velocity HI gas at -10 km/s with the legend in degrees K. (b) The same as (a) except overlaid on the MI contours at -120 km/s with the legend in degrees K. (c) IRAS 100 micron data showing the MI cavity with the legend in MJy/sr. (d) The diffuse soft X-ray image adapted from Fig. 2 of Herbstmeier et al. (1995) overlain on a map of the total HI column density between ±25 km/s with the legend in degrees K showing that the soft X-rays are found where the HI column density is lowest. The dashed circle highlights the approximate boundary of the low-velocity cavity seen in Fig. 1.